\documentclass{webofc}
\usepackage[varg]{txfonts}   
\usepackage{physics}
\usepackage[dvipsnames]{xcolor}
\usepackage{float}
\usepackage{siunitx}

\usepackage[varg]{txfonts}
\usepackage[capitalise]{cleveref}
\usepackage{ulem}

\begin{document}
\title{Impact of blending on weak lensing measurements with the Vera C. Rubin Observatory}
%
%

\author{\lastname{M. Ramel}\inst{1}\thanks{e-mail: manon.ramel@lpsc.in2p3.fr} \and
        \lastname{C. Doux}\inst{1} \and
        \lastname{M. Kuna}\inst{1} 
}

\institute{Université Grenoble Alpes, CNRS, LPSC-IN2P3, 38000 Grenoble, France
          }

\abstract{Upcoming deep optical surveys such as the Vera C. Rubin Observatory Legacy Survey of Space and Time will scan the sky to unprecedented depths and detect billions of galaxies. This amount of detections will however cause the apparent superposition of galaxies on the images, called blending, and generate a new systematic error due to the confusion of sources. As consequences, the measurements of individual galaxies properties such as their redshifts or shapes will be impacted, and some galaxies will not be detected. However, galaxy shapes are key quantities, used to estimate masses of large scale structures, such as galaxy clusters, through weak gravitational lensing. This work presents a new catalog matching algorithm, called \texttt{friendly}, for the detection and characterization of blends in simulated LSST data for the DESC Data Challenge 2. By identifying a specific type of blends, we show that removing them from the data may partially correct the amplitude of the $\Delta\Sigma$ weak lensing profile that could be biased low by around 20\% due to blending. This would result in impacting clusters weak lensing mass estimate and cosmology.}

\maketitle

\section{Introduction}
\label{sec1}

\subsection{Galaxy clusters for cosmology}
\label{subsec11}

Galaxy clusters are the largest gravitationally bound structures in the Universe. They are very massive and composed of 50 to 1000 of galaxies held together by gravity. Galaxy clusters trace the highest peaks in the total matter density field, including baryonic and dark matter. Since the number of matter overdensities and their masses are very dependent on the history and evolution of structure formation, clusters of galaxies are important astrophysical objects, used to infer cosmological parameters \cite{Allen2011Galaxyclusters}.

However, masses of galaxy clusters are not directly measurable and have to be determined through indirect effects such as weak gravitational lensing.
Gravitational lensing occurs in the presence of massive structures, such as galaxy clusters, acting as lenses between an observer and distant light sources, usually background galaxies \cite{Hoekstra2008weaklensing}. This configuration results in the deviation of light rays coming from the sources and passing close to the lens, and therefore in the distortion of background galaxy images. The measurement of the distorted shapes of source galaxies, $\epsilon^{\rm obs}$, allows to determine an estimator of the excess surface mass density $\widehat{\Delta\Sigma}$ \cite{Murray2022DeltaSigma} as function of the distance to the center of the structure, $R$, given as:

\begin{equation}
    \widehat{\Delta\Sigma}(R) = \expval{ \Sigma_{\rm crit}(z_{\rm gal}, z_l) \, \epsilon_+^{\rm obs} } (R),
\end{equation}
where ${\Sigma_{\rm crit}}$ is a geometrical term depending on the redshifts of the lens $z_l$ and of the source galaxies $z_{\rm{gal}}$, and $\epsilon_+^{\rm{obs}}$ are the tangential ellipticities of the sources. By fitting the $\Delta\Sigma$ profiles through a Navarro-Frenk-White halo mass profile model, estimated projected masses of galaxy clusters can be recovered~\cite{Navarro1996NFW}.

\subsection{The Vera C. Rubin Observatory}
\label{subsec12}

The next stage of future deep optical surveys will bring a large amount of data, including measurements of shapes and redshifts of galaxies, which will be used to estimate galaxy cluster masses through weak gravitational lensing. From 2025 onwards, the Legacy Survey of Space and Time (LSST) \cite{Ivezic2019LSST} survey will be conduct by the Vera C. Rubin Observatory which is currently in construction in northern Chile. During the ten years of observations of a \num{18000} squared degrees footprint, about 10 billions of galaxies up to a magnitude of 27 in $i$-band will be observed, representing one order of magnitude higher than previous optical surveys such as the Dark Energy Survey \cite{Abbott2016DES}.

\subsection{Blending}
\label{subsec13}

Due to the high depth of observation, the large number of observed galaxies and the effect of the atmosphere on future optical ground-based surveys such as LSST, an observational effect called blending will impact the measurement of the galaxy properties \cite{Dawson2015BlendingShapes}. Blending corresponds to the superposition of galaxies along the line of sight and on images, once they are projected onto 2D surfaces.

Two different types of blends can be distinguished. We define \textit{recognized blends} as two or more blended galaxies that overlap significantly but are detected as individual objects. This definition depends therefore on the detection pipeline. Deblenders such as \textsc{Scarlet} \cite{Melchior2018Scarlet} can be used to deblend recognized blended groups of galaxies, provided galaxy centers are properly identified. The fraction of recognized blends for LSST is estimated to be around 40~\%. However, some galaxies will be so overlapped that they will not even be identifiable as blends. These systems are called \textit{unrecognized blends}. Unrecognized blends are the most problematic ones since they will not be identifiable as such in future LSST data, by definition. Their proportion is estimated to be around 20~\% for an LSST-like survey \cite{Troxel2023BlendProportion} and, for the moment, no algorithm can correct for these blends. Therefore, blending will have an impact on the number of observed galaxies since several different galaxies will be identified as a unique object. Moreover, blending will impact the measurements of individual galaxies properties such as shapes or redshifts, and will bias weak lensing measurements induced by massive galaxy clusters.

\section{Matching procedure}
\label{sec2}

\subsection{Simulated catalogs}
\label{subsec21}

To study the impact of blending on future LSST weak lensing data, we first need to identify blended systems. To do so, it is necessary to compare LSST-like observation data with truth or reference data. This can be done by using simulations and matching two catalogs from the DESC collaboration \cite{Mandelbaum2018DESC}, as the LSST survey has not started yet.

The first one is \texttt{cosmoDC2} \cite{Korytov2019CosmoDC2}. It is a 440 squared degrees extragalactic catalog built from a dark matter N-body simulation and it is the starting point of the DESC Data Challenge 2 \cite{Abolfathi2021DC2}. Each galaxy in the catalog is characterized by properties such as their positions, true redshifts, intrinsic ellipticities or shears. They are also characterized by their dark matter haloes they were placed in and their Friends-of-Friends masses.

From the \texttt{cosmoDC2} catalog, 300 squared degrees have been simulated to create realistic images of the sky and processed by the Rubin science pipeline. Objects from these images have been detected and all their measured quantities, such as their positions, shapes and magnitudes, are contained in a second catalog called \texttt{DC2object}.
The comparison of these two catalogs allows to match the simulated truth with future observations from LSST, and therefore highlight a potential effect of blending between galaxies.


\subsection{\texttt{Friendly} algorithm}
\label{subsec22}
\subsubsection{Friends-of-Friends algorithm}
\label{subsubsec221}


To do such a comparison, we developed a new matching algorithm called \texttt{friendly}\footnote{https://github.com/LSSTDESC/friendly}, that better captures blends than existing algorithms. The first step of \texttt{friendly} is to use a Friends-of-Friends algorithm \footnote{https://github.com/yymao/FoFCatalogMatching} to identify groups of nearby objects and galaxies. This algorithm takes as input a distance in arcseconds called linking length, as well as the two catalogs one has to match. If two components of the catalogs are located within a radius equal to the linking length, they are linked to form a group. This algorithm allows to create networks, called groups, composed of simulated galaxies from \texttt{cosmoDC2} and corresponding detected objects from \texttt{DC2object}, based on their angular distances.

Once the groups have been formed, we can easily identify blended systems. In this paper, we will label $n-m$ systems those composed of $n$ galaxies and $m$ objects. The perfectly matched systems are composed of one true galaxy and one corresponding detected object and are therefore referenced as $1-1$ systems. The recognized blends are the $n-n$ systems, composed of the same number of objects and galaxies, this number being greater than one. Finally, the unrecognized blends are the $n-m$ systems with $n > m$ and $n,m>0$. We note immediately that the characteristics of those groups depends on the various cuts (e.g. on magnitude) applied to the two catalogs.

\subsubsection{Ellipse overlap test}
\label{subsubsec222}

The inconvenient of the Friends-of-Friends algorithm, is that it is only based on distances between points, corresponding to galaxies and objects, and therefore can be not sufficient for the study of extended blended sources. To resolve this issue, information about shapes of galaxies have been added to the Friends-of-Friends algorithm, leading to the development of the \texttt{friendly} matching tool.

To do so, we used an ellipse overlap test written by collaborator Shuang Liang\footnote{https://github.com/LSSTDESC/Cluster\_Blending/}. Based on the general ellipse equation parameters, this is binary test of overlap of the ellipses associated to galaxies (objects) based on true (measured) moments. Therefore, using this test, combined with the Friends-of-Friends algorithm with a relatively large linking length of 2~arcseconds, some groups can be refined using additional shapes information compared to only distances information.

\subsubsection{\texttt{NetworkX} graph structure}
\label{subsubsec223}

\begin{figure}
\centerline{\includegraphics[scale=0.74]{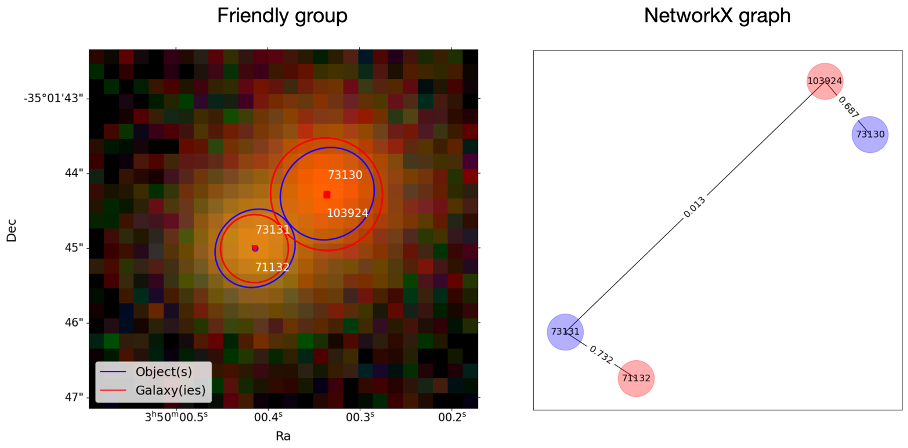}}
\caption{Left: example of one \texttt{friendly} group. Right: corresponding \texttt{NetworkX} graph. Ellipses and nodes associated to galaxies from \texttt{cosmoDC2} are in red, detected objects from \texttt{DC2object} are in blue. Absolute overlap fraction is indicated on each edge.}
\label{fig:friendly}
\end{figure}

In order to go further in the development of the \texttt{friendly} algorithm, a \texttt{NetworkX} graph structure has been implemented. This graph structure is composed of nodes, corresponding to galaxies or objects, linked through edges if they overlap, using the ellipse overlap test mentioned in \ref{subsubsec222}. An example of one \texttt{friendly} group and the corresponding associated \texttt{NetworkX} graph is shown in \cref{fig:friendly}. The graph structure is convenient to add metrics on the nodes and edges, and therefore add information about the constituents of blends. We implemented the absolute overlap fraction on each edge. It corresponds to the absolute surface of overlap between two different ellipses. Using a cut in this overlap fraction, some edges between nodes can be suppressed. Therefore, if two ellipses do not overlap with a significant surface, they can be considered as two isolated systems and some blended systems can be refined into smaller ones. The next step of this work is to add more metrics in the graphs such as the purity, colors, or magnitudes, among others, in order to refine blended systems and develop mitigation strategies.

\section{Impact of blending on $\Delta\Sigma$ profiles}
\label{sec3}

\begin{figure}
\centerline{\includegraphics[scale=0.35]{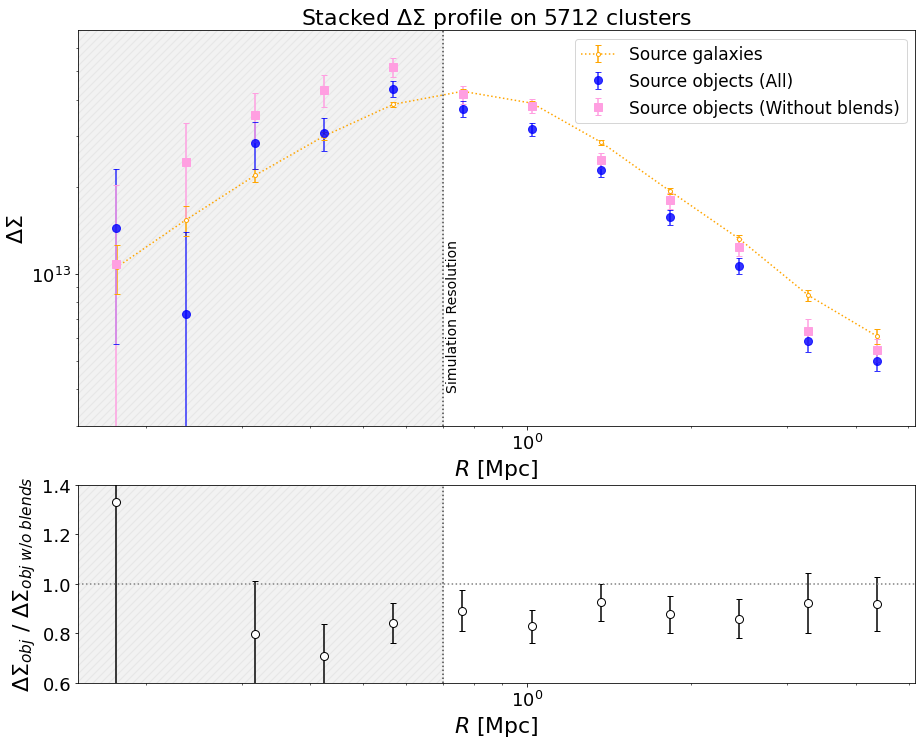}}
\caption{Top: impact of recognized blends on stacked $\Delta\Sigma$ profile. The yellow profile corresponds to the stacked profile measured using the shapes and redshifts of source galaxies from \texttt{cosmoDC2}. The blue (respectively pink) profile is measured with (respectively without) recognized blends from \texttt{DC2object} observation data. Bottom: ratio between the stacked profile measured on all detected source objects from \texttt{DC2object} and without the recognized blends.}
\label{fig:DeltaSigma}
\end{figure}

The objective of this work is to study the impact of both recognized and unrecognized blends on cluster $\Delta\Sigma$ profiles for the future LSST weak lensing data analysis. Only the results obtained in the case of recognized blends are shown in \cref{fig:DeltaSigma}. To do so, we plotted the stacked $\Delta\Sigma$ profiles of about \num{5000} galaxy clusters, since individual profiles are too noisy to be used.

In this proceeding, we present results that were first obtained by identifying the recognized blended systems using the Friends-of-Friends technique with a linking length of 0.4~arcseconds, corresponding to twice the pixel size. Then, we compared the stacked lensing profile measured using the shapes and redshifts of all detected objects from \texttt{DC2object} with the one obtained after removing recognized blends. For comparison purpose with simulated truth, we also plotted the stacked $\Delta\Sigma$ profile obtained by the lensed galaxies from \texttt{cosmoDC2}, which represents the profile of reference.

As a very preliminary result, we observe that removing blends may shift the lensing profile upwards by around 20 \%, so that it gets closer to the reference profile measured by \texttt{cosmoDC2} data. However, for future studies, removing recognized blends is not satisfactory since some cluster's galaxies can be suppressed from the data and therefore induce a bias in the estimated galaxy cluster mass. Moreover, the unrecognized blends will not be identifiable by definition, therefore no algorithm will be able to detect and suppress them from the future scientific data. This is why it is essential to deal with blends and mitigate their impact, by understanding their properties. This can be done, by using the in-development matching algorithm \texttt{friendly}. 

\section{Conclusion and perspectives}
\label{sec4}

In this work, we presented a new catalog matching algorithm, called \texttt{friendly}, developed to detect and characterize different types of blends in upcoming LSST weak lensing data. By combining a Friends-of-Friends algorithm, an ellipse overlap test and \texttt{NetworkX} graphs, we showed that \texttt{friendly} can be a robust tool for the study of blending and its impact on weak gravitational lensing profiles. As a prelimanary result, we found that recognized blends may induce a $\sim20\%$ bias in the amplitude of the stacked $\Delta\Sigma$ lensing profile, leading to a weaker signal and smaller galaxy clusters masses estimates. Further work is needed. In particular, the effect of well-defined \texttt{friendly} blended systems, separated into recognized and unrecognized blends, on weak lensing profiles needs to be studied. Finally, the impact of blending on estimated galaxy clusters masses and on cosmological parameters need to be propagated.

\end{document}